\documentclass[reprint, nofootinbib, amsmath,amssymb, aps, prl]{revtex4-2}
\usepackage[usenames,dvipsnames,table,xcdraw]{xcolor}
\usepackage[colorlinks,urlcolor=blue,citecolor=blue,linkcolor=blue, pdfborder={0 0 0}]{hyperref}
\usepackage{graphicx}
\usepackage{url}

\begin{document}

% \preprint{APS/zQED}

\title{The detectability of single spinless stellar-mass black holes through gravitational lensing of gravitational waves with advanced LIGO}
\author{Chengjiang Yin$^{1,2}$}
\author{Jian-hua He$^{1,2}$}
\thanks{Corresponding author: \href{mailto:hejianhua@nju.edu.cn}{hejianhua@nju.edu.cn}}
\affiliation{$^1$School of Astronomy and Space Science, Nanjing University, Nanjing 210093, P. R. China}
\affiliation{$^2$Key Laboratory of Modern Astronomy and Astrophysics (Nanjing University), Ministry of Education, Nanjing 210023, P. R. China}

\date{\today}

\begin{abstract}
We investigate the detectability of gravitational waves that have been lensed by a spinless stellar-mass black hole, with respect to the advanced LIGO. By solving the full relativistic linear wave equations in the spacetime of a Schwarzschild black hole, we find that the strong gravity can create unique signals in the lensed waveform, particularly during the merger and ringdown stages. The differences in terms of fitting factor between the lensed waveform and best-fitted unlensed general relativity template with spin-precessing and higher-order multipoles are greater than $5\%$ for the lens black hole mass within $70M_{\odot}<M_{\rm lens}<133.33 M_{\odot}$ under advanced LIGO's sensitivity. This is up to 5 times more detectable than the previous analysis based on the weak field approximation for a point mass and covers most part of the black hole mass gap predicted by stellar evolution theory. Based on Bayesian inference, the lensing feature can be distinguished with a signal-to-noise ratio of 12.5 for $M_{\rm lens}=70 M_{\odot}$ and 19.2 for $M_{\rm lens}=250 M_{\odot}$, which is attainable for advanced LIGO.
\end{abstract}

\maketitle

\textbf{Introduction}
Recent astronomical observations~\cite{liuWideStarBlackhole2019} and gravitational wave (GW) detections~\cite{LVC_GW190521_2020} have revealed the existence of massive stellar-mass black holes (BHs) within the mass gap~\cite{woosley_PI_2007}, which poses a potential challenge to the stellar evolution theory. While most of these BHs are observed in binary systems or under active accretion, the majority of stellar-mass BHs are believed to be single and quiescent~\cite{siciliaBlackHoleMass2022}. To comprehensively understand the formation of BHs in the mass gap, it is necessary to study single, inactive BHs. However, observing these BHs is an extreme challenge since they emit almost no detectable signals~\cite{Agol:2001hb}. Currently, the primary method to detect them is through gravitational microlensing~\cite{Bennett_2002,Mao,2022ApJ...933...83S}.

Similar to electromagnetic waves, GWs can also be lensed when passing through a BH. This provides a new probe to search for single stellar-mass BHs. Numerous studies have examined the lensed waveforms of BHs in the literature~\cite{Suyama:2005mx,PhysRevLett.80.1138,PhysRevD.98.083005,Ruffa_1999,DePaolis:2002tw,Zakharov_2002,Liao:2019aqq,Macquart:2004sh,Dai:2018enj,PhysRevD.90.062003,Christian,Meena:2019ate,2021PhRvD.103j4055W}. Nevertheless, most of them have been conducted under the weak field approximation and assumed a thin-lens model. This model neglects the impact of the strong gravity of BHs and breaks down when the incident wave is near the optical axis. For a point mass lens, the difference in terms of fitting factor between the lensed waveform and the best-fitted unlensed GR template is typically less than $1\%$, given the sensitivity of advanced LIGO (aLIGO)~\cite{PhysRevD.90.062003,PhysRevD.103.064047}. Even for the next generation GW detectors with much higher sensitivity, such as the LIGO Voyager~\cite{2020CQGra..37p5003A}, the Einstein Telescope~\cite{2011arXiv1108.1423S} and the LIGO Cosmic Explorer~\cite{2019BAAS...51g..35R}, the difference is no more than $4\%$~\cite{2021PhRvD.103j4055W}.
A recent search using data from GWTC-3 did not find any compelling evidence of lensing features predicted by the weak field approximation~\cite{LIGOScientific:2021izm, LIGOScientific:2023bwz}. It is worth noting that the microlensing results presented in the search should not be extended to scenarios involving strong gravity effects, such as BHs.

Indeed, unlike in the weak field limit or flat spacetime, GWs can interact with the background curvature and be scattered back in strong gravity field~\cite{1965ArRMA..18..103G,1990GReGr..22..843W,mclenaghan_1969,1992JMP....33..625S,1975weoc.book.....F},which can permanently change the waveform of GWs. For instance, when a finite wavelet with a clear trailing wavefront passes through a BH, typically the merger and ringdown signal of a binary coalescence, a long tail can emerge after the wavelet ~\cite{2022PhRvD.106l4037H}.
Despite its importance, this issue has not been addressed in the literature.

In addition to the strong gravity effect, the wave effect of GWs in the sensitive band of aLIGO (20 [Hz] $\sim$ 5000 [Hz])~\cite{LIGOScientific:2014pky} also becomes prominent for lens BHs with mass below $\sim 130M_{\odot}$\cite{Meena:2019ate}, where the GW wavelength is comparable to the event horizon radius of the lens BH. Unlike the geometric optics, GWs in aLIGO band can bypass their foreground stars and stellar-mass BHs along the optical axis due to diffraction~\cite{He:2021hhl}, while the electromagnetic radiation is sheltered by these dense opaque objects. Furthermore, in the thin-lens model, the frequency dependent lensing magnification factor derived from Kirchhoff diffraction integral diverges on the optic axis for a point mass lens~\cite{He:2021hhl}.

To fully address the above problems, we use a 3D time-domain numerical simulation, which solves the full relativistic linear perturbation equations in the spacetime of a Schwarzschild black hole. Throughout this letter, we adopt the geometric unit $c=G=1$, in which $1\ [\mathrm{Mpc}]=1.02938\times 10^{14}\ [\mathrm{Sec}]$ and $1 M_{\odot}=4.92535\times 10^{-6}\ [\mathrm{Sec}]$.

\begin{figure*}[htbp]
\begin{center}
\includegraphics[scale=0.85]{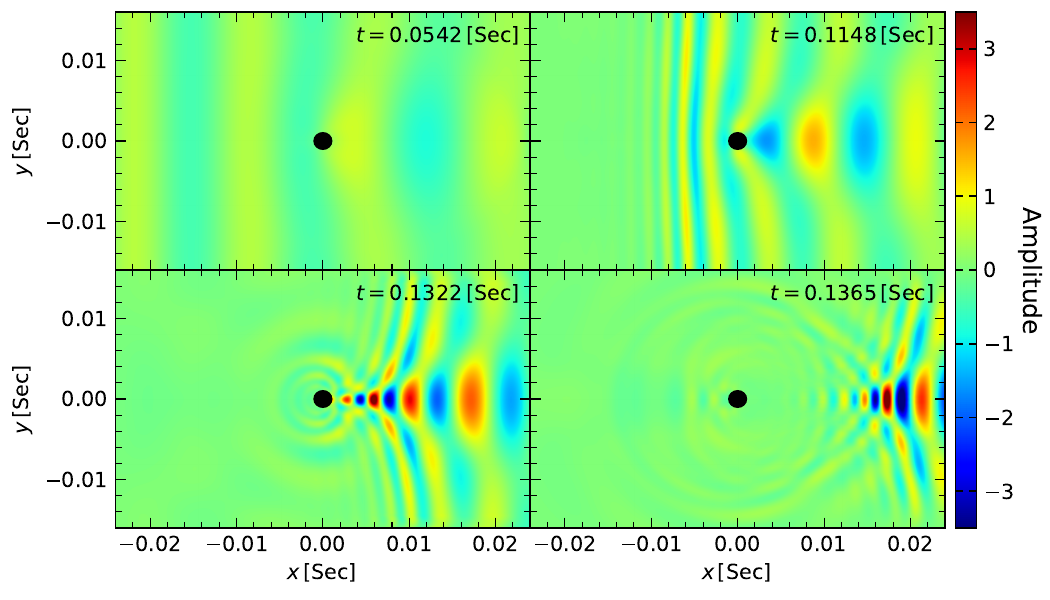}
\end{center}
\caption{Snapshots of our simulation at consecutive instants. They are taken along the $x-y$ plane at $z=0$.
The solid black circles indicate the position of the lens BH with radius equal to the BH event horizon. The colorbar to the right shows the amplitude of the GWs in our simulation.} \label{fig:snapshot}
\end{figure*}

\textbf{Numerical simulation}
The numerical setup is based on our previous work, as detailed in~\cite{2022PhRvD.106l4037H}. We use the \texttt{NRHybSur3dq8}~\cite{Varma:2018mmi}, which is a surrogate template model hybridized from non-precessing numerical relativity simulations, to generate the input waveform. We assume that the source is a non-spin quasi-circular binary BH system with equal masses and a total initial mass of 40$M_{\odot}$. The source is located 100.0 [Mpc] away from the lens and has an inclination of $\pi/2$, allowing for only $h_+$ polarization at the lens. The lensing effect primarily depends on the ratio between the incident GW wavelength and the event horizon radius of the lens BH $\rho_s$. This effect is not directly affected by spin-induced precession. We assume that the input waveform has zero spin and only $(2,\pm 2)$ modes. The input waveform lasts for 0.12933 [Sec], starting at 0.094666 [Sec] before the merger. As we shall show later, this waveform length is sufficient to obtain a complete lensed waveform through interpolation.

We assume that the lens is an isolated Schwarzschild BH with a mass of $133.33 M_{\odot}$. The shape of the simulation domain is a cylinder with a radius of $R_{\rm cy}= 48.73 \rho_s$ and a length of $L_{\rm cy}= 146.18 \rho_s$ ($x$-axis), where $\rho_s=M/2$ is the radius of the event horizon of the lens BH in isotropic coordinate. The cylinder length $L_{\rm cy}$ ensures that the gravity of the lens BH does not distort the input wavefront at the boundary of the simulation domain~\cite{2022PhRvD.106l4037H}. To get the lensed waveform, we place an observer at $x=0.023333\ [{\rm Sec}]\approx 73\rho_s$ on the $x$-axis after the BH. In practice, the simulation is performed with a scaling factor of $750$ and the degrees of freedom are $8.055\times 10^8$  using the first-order Lagrange element. The simulation uses 4800 CPU cores and consumes a total of 1.1M CPU-hours.

Figure~\ref{fig:snapshot} shows 4 snapshots of our simulation at various times, taken in the $x-y$ plane at $z=0$. The solid black circles represent the position of the lens BH with size equal to the event horizon. The color bar to the right shows the amplitude of the GW in our simulation.

After passing through the BH, GWs form a highly directional beam along the optical axis with an opening angle, which is related to the ratio between the GW wavelength of the incident waves and the radius of the event horizon of the lens BH $\rho_s$.This opening angle in our simulation is measured at a degree level, around $10^7$ times larger than the Einstein radius angle $\theta_E$, which is typically in the range of milliarcseconds for a stellar-mass BH in local galaxies~\cite{2022ApJ...933...83S}. Note that the lensing effect in the thin lens model of geometric optics becomes observable only when the source, lens, and observer are precisely aligned around $\theta_E$.  

\begin{figure*}[htbp]
\begin{center}
\includegraphics[scale=0.9]{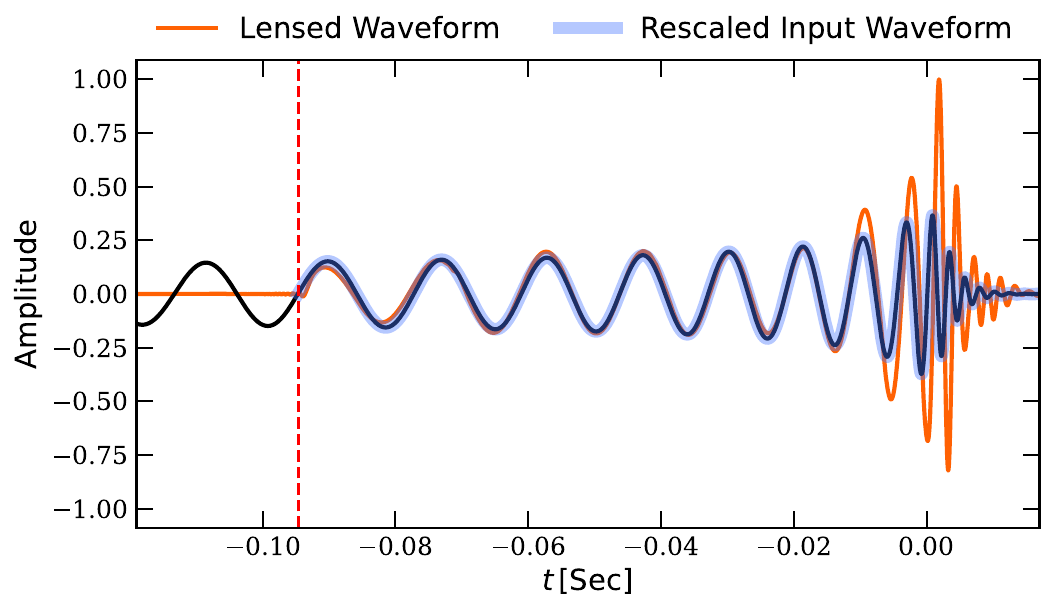}
\end{center}
\caption{The time domain lensed waveform (orange line) observed at $x=0.023333\ [{\rm Sec}]\approx 73\rho_s$ on the $x$-axis with amplitude normalized by its maximum value. The amplitude of the input waveform (blue shaded line) is also rescaled to match the lensed waveform during inspiral. For comparison, their starting time are aligned at the vertical dashed line with zero point time set at the merger of the input signal. The lensing effect affects the input waveform most significantly at the merger and ring-down stages while the inspiral signals match well with each other after rescaling. The lensed waveform can thus be interpolated down to $f_{\rm lower} = 20\ [\mathrm{Hz}]$ using the rescaled input waveform (black solid line).}\label{fig:waveform}
\end{figure*}

Figure~\ref{fig:waveform} shows the lensed waveform (orange line) observed at $x=0.023333\ [{\rm Sec}]\approx 73\rho_s$ on the $x$-axis with amplitude normalized by its maximum value. Since the lensed waveform is only amplified during inspiral, we rescale the input waveform (blue shaded line) to match the lensed waveform before -0.04 [Sec] for better comparison. While in the merger and ring-down stages, the lensed waveform undergoes significant frequency-dependent amplification and phase shift.

\textbf{The detectability with respect to aLIGO}
We first explore the potential degeneracy between the lensing effect and known physical effects from the unlensed GR templates. We do not consider templates that are beyond GR in this work. We choose the fully spin-precessing time-domain effective-one-body waveform model \texttt{SEOBNRv4PHM}~\cite{2020PhRvD.102d4055O} to generate the unlensed GR waveforms. High order multipoles are considered up to $l\leq 4$. The free parameter for the unlensed GR template is $(M_{\rm total},q_s,\chi_1,\theta_1,\phi_1,\chi_2,\theta_2,\phi_2,\iota,t_0,\phi_0)$. Here $M_{\rm total}$ is the total initial binary mass. $q_s$ is the mass ratio of the source binary. $\chi_{1,2}$ is the dimensionless spin. $\theta_{1,2}$ and $\phi_{1,2}$ are zenith and azimuthal angles between spin and the Newtonian orbital angular momentum $L_N$. $\iota$ is the inclination angle of $L_N$. $t_0$ and $\phi_0$ is the time and phase offset.

We use the SNR in matched filter to quantitatively evaluate the differences between the lensed and unlensed GR templates. We first define a complex number correlation
\begin{align}
    z(t_0)
    &=\frac{4}{\sigma(h_{\rm lensed}) \cdot \sigma(h_{\rm GR, 0})}\nonumber\\ 
    &\times \int_0^{\infty} \frac{\tilde{h}_{\rm lensed}(f)\cdot \tilde{h}_{\rm GR, 0}^*(f)}{S_n(f)}e^{2\pi i f t_0}\,\mathrm{d}f \,,
    \label{z_complex}
\end{align}
where $h_{\rm lensed}$ is the lensed waveform and $h_{\rm GR, 0}$ is the unlensed GR template with unoptimized time and phase offset, \textit{i.e.} $t_0=0$, $\phi_0=0$. $S_n(f)$ is the power spectrum density (PSD) of the detector. $\sigma(\cdot)$ is a normalization constant for the given waveform defined as
\begin{equation}
    \sigma^2(h)=4\int_0^{\infty} \frac{|\tilde{h}(f)|^2}{S_n(f)}\, \mathrm{d}f\,.
\end{equation}

The fitting factor ($\mathrm{FF}$) is defined as the maximum modulus of the complex correlation \cite{PhysRevD.52.605}
\begin{equation}
\mathrm{FF}= \max_{t_0}\,|z(t_0)|\,.
\end{equation}
The time offset $t_0$ is optimized when it gives out $\mathrm{FF}$ and the corresponding phase offset $2\phi_{\rm 0}={\rm arg}\,z(t_0)$.

In this work, we choose $S_n(f)$ the sensitivity curve of the aLIGO detector at Hanford based on the first three months during O3\footnote{\url{https://dcc.ligo.org/LIGO-T2000012/public}}. We assume that the length of the signal is $T=512$ [Sec], leading to a PSD resolution of $\Delta f = 1/T=0.001953125$ [Hz]. The lower bound of the waveform frequency is set as $f_{\rm lower} = 20\ [\mathrm{Hz}]$. Given the fact that the rescaled input waveform matches the lensed waveform very well during inspiral, we interpolate the lensed waveform down to $f_{\rm lower}$ using the rescaled input waveform (black solid line in Fig.~\ref{fig:waveform}). 

\begin{figure*}[htbp]
\begin{center}
\includegraphics[scale=0.9]{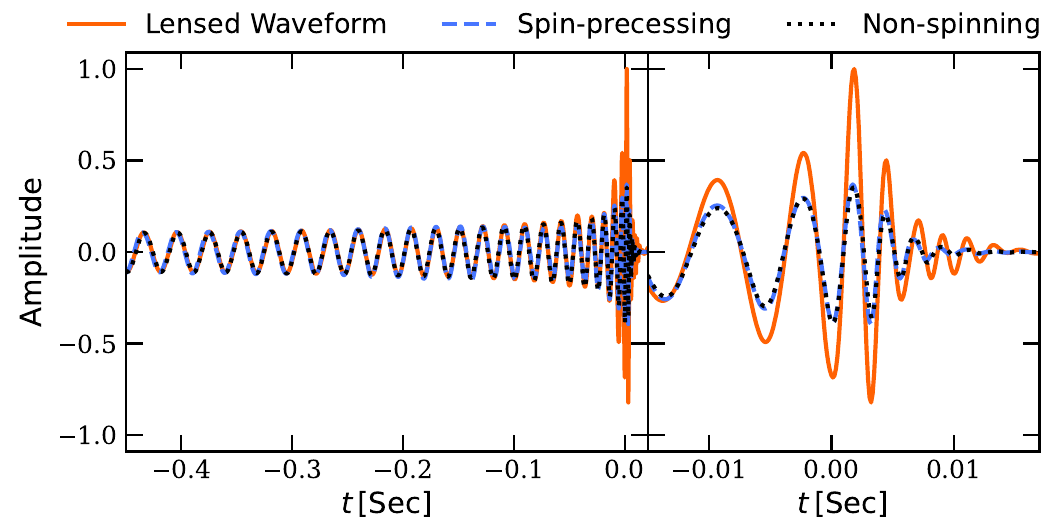}
\end{center}
\caption{Comparison between the lensed GW waveform and the best-fitted unlensed GR template. The solid orange line is for the lensed GW waveform. The dashed blue line and dotted black line show the best-fitted unlensed GR template with and without spin-precessing respectively. The left panel shows the overall waveforms and the right panel shows a zoom-in view at the merger and ringdown stages.}
\label{fig:best-fitted}
\end{figure*}

Since $t_0$ and $\phi_0$ can be efficiently calculated and optimized from Eq.~(\ref{z_complex}) via the fast Fourier transform, we focus on the remaining 9 parameters and use the Markov Chain Monte Carlo (MCMC) to explore the parameter space based on the \texttt{PyCBC} pipeline~\cite{Biwer:2018osg}. We set the priors for the parameters as $M_{\rm total}\in[38 M_{\odot},42 M_{\odot}]\,$, $q_s\in[0.45,1]\,$, $\theta_{1,2}\in[0,\pi]\,$, $\phi_{1,2}\in[0,2\pi]\,$, $\iota\in[0,\pi/2]\,$. The priors for the dimensionless spin is set as $\chi_{1,2}\in[0,0.7]\,$, within which the precessing accuracy of \texttt{SEOBNRv4PHM} in terms of the fitting factor is better than $1\%$ against the numerical relativity simulation in most cases~\cite{2022Natur.610..652H}. Considering that the waveform typically only lasts for around 2 [Sec] for our parameter space, given the lower frequency cut-off $f_{\rm lower} = 20$ [Hz], we zero-pad the length to 8 [Sec], as is commonly done in \texttt{PyCBC}. We sample the parameter space with 50 walkers and 1000 iterations for each. We then find the fitting factor among these 50000 samples. The best-fitted waveforms are shown in Fig.~\ref{fig:best-fitted}. For the spin-precessing templates, we find $\mathrm{FF}_\mathrm{spin}=0.946$. And for non-spin templates, namely $\chi_{1,2}=0$, we find $\mathrm{FF}_\mathrm{non- spin}=0.926$. This demonstrates that spin-precessing partially degenerates with the lensing effect.

To generate the lensed waveforms for different values of the lens mass, we rescale our numerical results, which is a valid approach since we have used the geometric unit and our wave equations are fully linear~\cite{2022PhRvD.106l4037H}. The rescaling factor is between $0.5$ and $1.875$, which corresponds to a total source binary mass range of $20 M_{\odot}$ to $75 M_{\odot}$. This covers the majority of events detected by aLIGO. The rescaling keeps the ratio between the incident GW wavelength at coalescence, $\lambda_\mathrm{coal}$, and the event horizon radius of the lens BH, $\rho_s$, around 9.7, where the lens BH has relatively significant effect on the lensed waveform. For each rescaled lensed waveform, we use the same pipeline to find the fitting factors.

\begin{figure}[htbp]
\begin{center}
\includegraphics[width=1\columnwidth]{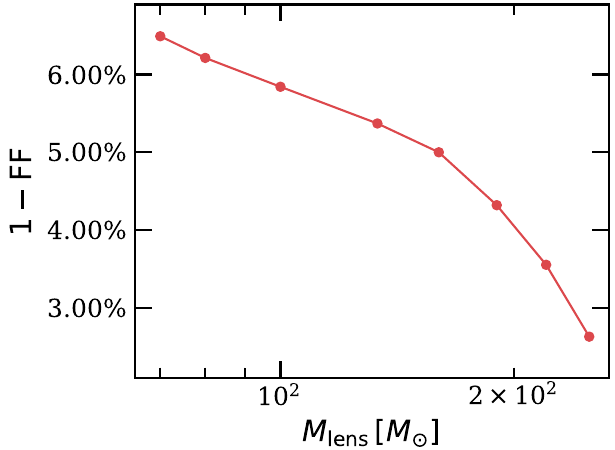}
\end{center}
\caption{The differences ($1-\mathrm{ FF}$) between the lensed waveforms and the best-fitted unlensed GR templates as a function of the lens BH mass. The unlensed GR templates include spin-precessing and higher order modes. The fitting factors are obtained by exploring the full parameter space of the GR templates using MCMC.}
\label{fig:Mismatch}
\end{figure}

Figure~\ref{fig:Mismatch} shows the differences ($1-\mathrm{ FF}$) between the lensed waveforms and the best-fitted spin-precessing templates as a function of the lens mass. For $M_{\rm lens}<133.33 M_{\odot}$ even with spin-precessing, the differences between the lensed waveform and the unlensed GR templates are substantial, with ($1-\mathrm{ FF}$) greater than $5\%$.

We use the Bayesian inference to estimate whether such a lensing signal can be distinguished by aLIGO.
When ($1-\mathrm{FF}$) is not too large, the Bayes factor between the lensed and unlensed GR template is given by~\cite{Cornish:2011ys,DelPozzo:2014cla}
\begin{align}
\ln B_{\rm lensed, unlensed}
&\approx\frac{\mathrm{SNR}^2}{2}(1-\mathrm{ FF}^2)\nonumber\\
&+(n_{\rm GR}-\delta n_{\rm lensed})\ln \mathrm{FF} \,.
\end{align}

Here $n_{\rm GR}=15$ is the total number of parameters in the unlensed GR templates including external parameters. $\delta n_{\rm lensed}=1$ is the number of additional parameters in the lensed waveform relative to the unlensed case.

\begin{figure}[htbp]
\begin{center}
\includegraphics[width=1\columnwidth]{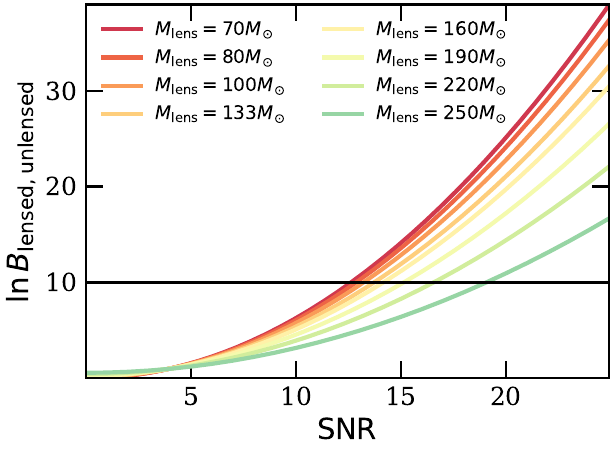}
\end{center}
\caption{The logarithm Bayes factor between the lensed waveform and the best-fitted unlensed GR templates as a function of the SNR under aLIGO O3's sensitivity. The solid black horizontal line indicates the detecting threshold of $\ln B_{\rm lensed, unlensed} =10$.}
\label{fig:bayes}
\end{figure}

Figure~\ref{fig:bayes} shows the logarithm Bayes factor between the lensed waveform and the best-fitted unlensed GR templates as a function of the SNR under aLIGO O3's sensitivity. The solid black horizontal line indicates the detecting threshold of $\ln B_{\rm lensed, unlensed} =10$. Given this criterion, for $M_{\rm lens}=70 M_{\odot}$ the lensed waveform can be distinguished from the unlensed ones if the SNR is greater than 12.5. For $M_{\rm lens}=250 M_{\odot}$, the required SNR is $19.2$, which is attainable for aLIGO.

\textbf{Discussions}
When GWs pass through a BH with the event horizon comparable to the GW wavelength, gravitational lensing will occur and forms a strongly directional beam along the optical axis due to the wave nature of GWs.
The lensing effect can be detected as long as observed within the beam opening angle, which is at a degree level and is independent to the distances of source and lens.This opening angle is approximately $10^7$ times larger than the Einstein radius angle $\theta_E$ in the thin lens model of geometric optics for a stellar-mass BH in local galaxies. Additionally, the beam is not only amplified but also has a complicated non-spherical wavefront and therefore decays slower than spherical waves.
All of these can enhance the detection probability of lensing events.

The waveform within the lensing beam also undergoes considerable frequency-dependent amplification and phase shift due to the strong gravity of the lens BH, particularly during the merger and ringdown stages.
This unique lensing signal can not be mimicked by the unlensed GR template even with spin-precessing and can be distinguished under aLIGO's sensitivity with moderate SNR, which is more detectable than the prediction based on weak field approximation.

On the other hand, we can also use the lensed waveform to deduce the physical properties of the lens BH.
This offers a new way to detect single and quiescent BHs, which are believed to account for a dominant fraction of stellar-mass BHs.
Given substantial detections of these solitary objects, a more comprehensive understanding of the BH distribution can be achieved. This can potentially help to reduce the uncertainty of the lower bound of the BH mass gap. Besides, this method can also be used in the search for primordial BHs since they are considered to be isolated and spinless~\cite{PBH_spin_2017}.

Finally, to actually search for such lensing signals in LIGO-Virgo observations or even the next-generation GW detectors, a template bank with gravitational lensing effect should be established. To achieve this, we will use phenomenological models with an ansatz in the frequency domain. These models will be calibrated in the time domain against a carefully selected set of simulations, which encompass various parameter sets, spanning a wide range of ratios between the GW wavelength and the lens BH horizon radius. A detailed analysis will be provided in our following work.

\vspace{5mm}
\noindent \textbf{Acknowledgments}
This work is supported by the National Key R$\&$D Program of China (Grant No. 2021YFC2203002), the National Natural Science Foundation of China (Grants No. 12075116, No. 12150011), the science research grants from the China Manned Space Project (Grant No. CMS-CSST-2021-A03), the Natural Science Foundation of the Jiangsu Higher Education Institutions of China (Grant No. 22KJB630006), the Fundamental Research Funds for the Central Universities (Grant No. 020114380053).
The numerical calculations in this letter have been done on the computing facilities in the High Performance Computing Center (HPCC) of Nanjing University.

\bibliography{bibliography}

\begin{thebibliography}{43}
\expandafter\ifx\csname natexlab\endcsname\relax\def\natexlab#1{#1}\fi
\expandafter\ifx\csname bibnamefont\endcsname\relax
  \def\bibnamefont#1{#1}\fi
\expandafter\ifx\csname bibfnamefont\endcsname\relax
  \def\bibfnamefont#1{#1}\fi
\expandafter\ifx\csname citenamefont\endcsname\relax
  \def\citenamefont#1{#1}\fi
\expandafter\ifx\csname url\endcsname\relax
  \def\url#1{\texttt{#1}}\fi
\expandafter\ifx\csname urlprefix\endcsname\relax\def\urlprefix{URL }\fi
\providecommand{\bibinfo}[2]{#2}
\providecommand{\eprint}[2][]{\url{#2}}

\bibitem[{\citenamefont{Liu et~al.}(2019)\citenamefont{Liu, Zhang, Howard, Bai,
  Lu, Soria, Justham, Li, Zheng, Wang et~al.}}]{liuWideStarBlackhole2019}
\bibinfo{author}{\bibfnamefont{J.}~\bibnamefont{Liu}},
  \bibinfo{author}{\bibfnamefont{H.}~\bibnamefont{Zhang}},
  \bibinfo{author}{\bibfnamefont{A.~W.} \bibnamefont{Howard}},
  \bibinfo{author}{\bibfnamefont{Z.}~\bibnamefont{Bai}},
  \bibinfo{author}{\bibfnamefont{Y.}~\bibnamefont{Lu}},
  \bibinfo{author}{\bibfnamefont{R.}~\bibnamefont{Soria}},
  \bibinfo{author}{\bibfnamefont{S.}~\bibnamefont{Justham}},
  \bibinfo{author}{\bibfnamefont{X.}~\bibnamefont{Li}},
  \bibinfo{author}{\bibfnamefont{Z.}~\bibnamefont{Zheng}},
  \bibinfo{author}{\bibfnamefont{T.}~\bibnamefont{Wang}}, \bibnamefont{et~al.},
  \bibinfo{journal}{Nature} \textbf{\bibinfo{volume}{575}},
  \bibinfo{pages}{618} (\bibinfo{year}{2019}), ISSN \bibinfo{issn}{0028-0836,
  1476-4687}.

\bibitem[{\citenamefont{Abbott et~al.}(2020)}]{LVC_GW190521_2020}
\bibinfo{author}{\bibfnamefont{R.}~\bibnamefont{Abbott}} \bibnamefont{et~al.}
  (\bibinfo{collaboration}{LIGO Scientific Collaboration and Virgo
  Collaboration}), \bibinfo{journal}{Astrophys. J. Lett.}
  \textbf{\bibinfo{volume}{900}}, \bibinfo{pages}{L13} (\bibinfo{year}{2020}),
  ISSN \bibinfo{issn}{2041-8205}.

\bibitem[{\citenamefont{{Woosley} et~al.}(2007)\citenamefont{{Woosley},
  {Blinnikov}, and {Heger}}}]{woosley_PI_2007}
\bibinfo{author}{\bibfnamefont{S.~E.} \bibnamefont{{Woosley}}},
  \bibinfo{author}{\bibfnamefont{S.}~\bibnamefont{{Blinnikov}}},
  \bibnamefont{and} \bibinfo{author}{\bibfnamefont{A.}~\bibnamefont{{Heger}}},
  \bibinfo{journal}{Nature} \textbf{\bibinfo{volume}{450}},
  \bibinfo{pages}{390} (\bibinfo{year}{2007}), \eprint{0710.3314}.

\bibitem[{\citenamefont{Sicilia et~al.}(2022)\citenamefont{Sicilia, Lapi, Boco,
  Spera, Carlo, Mapelli, Shankar, Alexander, Bressan, and
  Danese}}]{siciliaBlackHoleMass2022}
\bibinfo{author}{\bibfnamefont{A.}~\bibnamefont{Sicilia}},
  \bibinfo{author}{\bibfnamefont{A.}~\bibnamefont{Lapi}},
  \bibinfo{author}{\bibfnamefont{L.}~\bibnamefont{Boco}},
  \bibinfo{author}{\bibfnamefont{M.}~\bibnamefont{Spera}},
  \bibinfo{author}{\bibfnamefont{U.~N.~D.} \bibnamefont{Carlo}},
  \bibinfo{author}{\bibfnamefont{M.}~\bibnamefont{Mapelli}},
  \bibinfo{author}{\bibfnamefont{F.}~\bibnamefont{Shankar}},
  \bibinfo{author}{\bibfnamefont{D.~M.} \bibnamefont{Alexander}},
  \bibinfo{author}{\bibfnamefont{A.}~\bibnamefont{Bressan}}, \bibnamefont{and}
  \bibinfo{author}{\bibfnamefont{L.}~\bibnamefont{Danese}},
  \bibinfo{journal}{Astrophys. J.} \textbf{\bibinfo{volume}{924}},
  \bibinfo{pages}{56} (\bibinfo{year}{2022}).

\bibitem[{\citenamefont{{Agol} and {Kamionkowski}}(2002)}]{Agol:2001hb}
\bibinfo{author}{\bibfnamefont{E.}~\bibnamefont{{Agol}}} \bibnamefont{and}
  \bibinfo{author}{\bibfnamefont{M.}~\bibnamefont{{Kamionkowski}}},
  \bibinfo{journal}{Mon. Not. Roy. Astron. Soc.}
  \textbf{\bibinfo{volume}{334}}, \bibinfo{pages}{553} (\bibinfo{year}{2002}),
  \eprint{astro-ph/0109539}.

\bibitem[{\citenamefont{Bennett et~al.}(2002)\citenamefont{Bennett, Becker,
  Quinn, Tomaney, Alcock, Allsman, Alves, Axelrod, Calitz, Cook
  et~al.}}]{Bennett_2002}
\bibinfo{author}{\bibfnamefont{D.~P.} \bibnamefont{Bennett}},
  \bibinfo{author}{\bibfnamefont{A.~C.} \bibnamefont{Becker}},
  \bibinfo{author}{\bibfnamefont{J.~L.} \bibnamefont{Quinn}},
  \bibinfo{author}{\bibfnamefont{A.~B.} \bibnamefont{Tomaney}},
  \bibinfo{author}{\bibfnamefont{C.}~\bibnamefont{Alcock}},
  \bibinfo{author}{\bibfnamefont{R.~A.} \bibnamefont{Allsman}},
  \bibinfo{author}{\bibfnamefont{D.~R.} \bibnamefont{Alves}},
  \bibinfo{author}{\bibfnamefont{T.~S.} \bibnamefont{Axelrod}},
  \bibinfo{author}{\bibfnamefont{J.~J.} \bibnamefont{Calitz}},
  \bibinfo{author}{\bibfnamefont{K.~H.} \bibnamefont{Cook}},
  \bibnamefont{et~al.}, \bibinfo{journal}{Astrophys. J.}
  \textbf{\bibinfo{volume}{579}}, \bibinfo{pages}{639} (\bibinfo{year}{2002}).

\bibitem[{\citenamefont{Mao et~al.}(2002)\citenamefont{Mao, Smith, Woźniak,
  Udalski, Szymański, Kubiak, Pietrzyński, Soszyński, and Żebruń}}]{Mao}
\bibinfo{author}{\bibfnamefont{S.}~\bibnamefont{Mao}},
  \bibinfo{author}{\bibfnamefont{M.~C.} \bibnamefont{Smith}},
  \bibinfo{author}{\bibfnamefont{P.}~\bibnamefont{Woźniak}},
  \bibinfo{author}{\bibfnamefont{A.}~\bibnamefont{Udalski}},
  \bibinfo{author}{\bibfnamefont{M.}~\bibnamefont{Szymański}},
  \bibinfo{author}{\bibfnamefont{M.}~\bibnamefont{Kubiak}},
  \bibinfo{author}{\bibfnamefont{G.}~\bibnamefont{Pietrzyński}},
  \bibinfo{author}{\bibfnamefont{I.}~\bibnamefont{Soszyński}},
  \bibnamefont{and} \bibinfo{author}{\bibfnamefont{K.}~\bibnamefont{Żebruń}},
  \bibinfo{journal}{Mon. Not. Roy. Astron. Soc.}
  \textbf{\bibinfo{volume}{329}}, \bibinfo{pages}{349} (\bibinfo{year}{2002}),
  ISSN \bibinfo{issn}{0035-8711}.

\bibitem[{\citenamefont{{Sahu} et~al.}(2022)\citenamefont{{Sahu}, {Anderson},
  {Casertano}, {Bond}, {Udalski}, {Dominik}, {Calamida}, {Bellini}, {Brown},
  {Rejkuba} et~al.}}]{2022ApJ...933...83S}
\bibinfo{author}{\bibfnamefont{K.~C.} \bibnamefont{{Sahu}}},
  \bibinfo{author}{\bibfnamefont{J.}~\bibnamefont{{Anderson}}},
  \bibinfo{author}{\bibfnamefont{S.}~\bibnamefont{{Casertano}}},
  \bibinfo{author}{\bibfnamefont{H.~E.} \bibnamefont{{Bond}}},
  \bibinfo{author}{\bibfnamefont{A.}~\bibnamefont{{Udalski}}},
  \bibinfo{author}{\bibfnamefont{M.}~\bibnamefont{{Dominik}}},
  \bibinfo{author}{\bibfnamefont{A.}~\bibnamefont{{Calamida}}},
  \bibinfo{author}{\bibfnamefont{A.}~\bibnamefont{{Bellini}}},
  \bibinfo{author}{\bibfnamefont{T.~M.} \bibnamefont{{Brown}}},
  \bibinfo{author}{\bibfnamefont{M.}~\bibnamefont{{Rejkuba}}},
  \bibnamefont{et~al.}, \bibinfo{journal}{Astrophys. J}
  \textbf{\bibinfo{volume}{933}}, \bibinfo{eid}{83} (\bibinfo{year}{2022}),
  \eprint{2201.13296}.

\bibitem[{\citenamefont{Suyama et~al.}(2005)\citenamefont{Suyama, Takahashi,
  and Michikoshi}}]{Suyama:2005mx}
\bibinfo{author}{\bibfnamefont{T.}~\bibnamefont{Suyama}},
  \bibinfo{author}{\bibfnamefont{R.}~\bibnamefont{Takahashi}},
  \bibnamefont{and}
  \bibinfo{author}{\bibfnamefont{S.}~\bibnamefont{Michikoshi}},
  \bibinfo{journal}{Phys. Rev. D} \textbf{\bibinfo{volume}{72}},
  \bibinfo{pages}{043001} (\bibinfo{year}{2005}), \eprint{astro-ph/0505023}.

\bibitem[{\citenamefont{Nakamura}(1998)}]{PhysRevLett.80.1138}
\bibinfo{author}{\bibfnamefont{T.~T.} \bibnamefont{Nakamura}},
  \bibinfo{journal}{Phys. Rev. Lett.} \textbf{\bibinfo{volume}{80}},
  \bibinfo{pages}{1138} (\bibinfo{year}{1998}).

\bibitem[{\citenamefont{Lai et~al.}(2018)\citenamefont{Lai, Hannuksela,
  Herrera-Mart\'{\i}n, Diego, Broadhurst, and Li}}]{PhysRevD.98.083005}
\bibinfo{author}{\bibfnamefont{K.-H.} \bibnamefont{Lai}},
  \bibinfo{author}{\bibfnamefont{O.~A.} \bibnamefont{Hannuksela}},
  \bibinfo{author}{\bibfnamefont{A.}~\bibnamefont{Herrera-Mart\'{\i}n}},
  \bibinfo{author}{\bibfnamefont{J.~M.} \bibnamefont{Diego}},
  \bibinfo{author}{\bibfnamefont{T.}~\bibnamefont{Broadhurst}},
  \bibnamefont{and} \bibinfo{author}{\bibfnamefont{T.~G.~F.} \bibnamefont{Li}},
  \bibinfo{journal}{Phys. Rev. D} \textbf{\bibinfo{volume}{98}},
  \bibinfo{pages}{083005} (\bibinfo{year}{2018}),
  \urlprefix\url{https://link.aps.org/doi/10.1103/PhysRevD.98.083005}.

\bibitem[{\citenamefont{Ruffa}(1999)}]{Ruffa_1999}
\bibinfo{author}{\bibfnamefont{A.~A.} \bibnamefont{Ruffa}},
  \bibinfo{journal}{Astrophys. J.} \textbf{\bibinfo{volume}{517}},
  \bibinfo{pages}{L31} (\bibinfo{year}{1999}),
  \urlprefix\url{https://doi.org/10.1086%2F312015}.

\bibitem[{\citenamefont{De~Paolis et~al.}(2002)\citenamefont{De~Paolis,
  Ingrosso, Nucita, and Qadir}}]{DePaolis:2002tw}
\bibinfo{author}{\bibfnamefont{F.}~\bibnamefont{De~Paolis}},
  \bibinfo{author}{\bibfnamefont{G.}~\bibnamefont{Ingrosso}},
  \bibinfo{author}{\bibfnamefont{A.~A.} \bibnamefont{Nucita}},
  \bibnamefont{and} \bibinfo{author}{\bibfnamefont{A.}~\bibnamefont{Qadir}},
  \bibinfo{journal}{Astron. Astrophys.} \textbf{\bibinfo{volume}{394}},
  \bibinfo{pages}{749} (\bibinfo{year}{2002}), \eprint{astro-ph/0209149}.

\bibitem[{\citenamefont{Zakharov and Baryshev}(2002)}]{Zakharov_2002}
\bibinfo{author}{\bibfnamefont{A.~F.} \bibnamefont{Zakharov}} \bibnamefont{and}
  \bibinfo{author}{\bibfnamefont{Y.~V.} \bibnamefont{Baryshev}},
  \bibinfo{journal}{Classical and Quantum Gravity}
  \textbf{\bibinfo{volume}{19}}, \bibinfo{pages}{1361} (\bibinfo{year}{2002}).

\bibitem[{\citenamefont{Liao et~al.}(2019)\citenamefont{Liao, Biesiada, and
  Fan}}]{Liao:2019aqq}
\bibinfo{author}{\bibfnamefont{K.}~\bibnamefont{Liao}},
  \bibinfo{author}{\bibfnamefont{M.}~\bibnamefont{Biesiada}}, \bibnamefont{and}
  \bibinfo{author}{\bibfnamefont{X.-L.} \bibnamefont{Fan}},
  \bibinfo{journal}{Astrophys. J.} \textbf{\bibinfo{volume}{875}},
  \bibinfo{pages}{139} (\bibinfo{year}{2019}), \eprint{1903.06612}.

\bibitem[{\citenamefont{Macquart}(2004)}]{Macquart:2004sh}
\bibinfo{author}{\bibfnamefont{J.-P.} \bibnamefont{Macquart}},
  \bibinfo{journal}{Astron. Astrophys.} \textbf{\bibinfo{volume}{422}},
  \bibinfo{pages}{761} (\bibinfo{year}{2004}), \eprint{astro-ph/0402661}.

\bibitem[{\citenamefont{Dai et~al.}(2018)\citenamefont{Dai, Li, Zackay, Mao,
  and Lu}}]{Dai:2018enj}
\bibinfo{author}{\bibfnamefont{L.}~\bibnamefont{Dai}},
  \bibinfo{author}{\bibfnamefont{S.-S.} \bibnamefont{Li}},
  \bibinfo{author}{\bibfnamefont{B.}~\bibnamefont{Zackay}},
  \bibinfo{author}{\bibfnamefont{S.}~\bibnamefont{Mao}}, \bibnamefont{and}
  \bibinfo{author}{\bibfnamefont{Y.}~\bibnamefont{Lu}}, \bibinfo{journal}{Phys.
  Rev. D} \textbf{\bibinfo{volume}{98}}, \bibinfo{pages}{104029}
  (\bibinfo{year}{2018}), \eprint{1810.00003}.

\bibitem[{\citenamefont{Cao et~al.}(2014)\citenamefont{Cao, Li, and
  Wang}}]{PhysRevD.90.062003}
\bibinfo{author}{\bibfnamefont{Z.}~\bibnamefont{Cao}},
  \bibinfo{author}{\bibfnamefont{L.-F.} \bibnamefont{Li}}, \bibnamefont{and}
  \bibinfo{author}{\bibfnamefont{Y.}~\bibnamefont{Wang}},
  \bibinfo{journal}{Phys. Rev. D} \textbf{\bibinfo{volume}{90}},
  \bibinfo{pages}{062003} (\bibinfo{year}{2014}),
  \urlprefix\url{https://link.aps.org/doi/10.1103/PhysRevD.90.062003}.

\bibitem[{\citenamefont{Christian et~al.}(2018)\citenamefont{Christian, Vitale,
  and Loeb}}]{Christian}
\bibinfo{author}{\bibfnamefont{P.}~\bibnamefont{Christian}},
  \bibinfo{author}{\bibfnamefont{S.}~\bibnamefont{Vitale}}, \bibnamefont{and}
  \bibinfo{author}{\bibfnamefont{A.}~\bibnamefont{Loeb}},
  \bibinfo{journal}{Phys. Rev. D} \textbf{\bibinfo{volume}{98}},
  \bibinfo{pages}{103022} (\bibinfo{year}{2018}),
  \urlprefix\url{https://link.aps.org/doi/10.1103/PhysRevD.98.103022}.

\bibitem[{\citenamefont{{Meena} and {Bagla}}(2020)}]{Meena:2019ate}
\bibinfo{author}{\bibfnamefont{A.~K.} \bibnamefont{{Meena}}} \bibnamefont{and}
  \bibinfo{author}{\bibfnamefont{J.~S.} \bibnamefont{{Bagla}}},
  \bibinfo{journal}{Mon. Not. Roy. Astron. Soc.}
  \textbf{\bibinfo{volume}{492}}, \bibinfo{pages}{1127} (\bibinfo{year}{2020}),
  \eprint{1903.11809}.

\bibitem[{\citenamefont{{Wang} et~al.}(2021)\citenamefont{{Wang}, {Lo}, {Li},
  and {Chen}}}]{2021PhRvD.103j4055W}
\bibinfo{author}{\bibfnamefont{Y.}~\bibnamefont{{Wang}}},
  \bibinfo{author}{\bibfnamefont{R.~K.~L.} \bibnamefont{{Lo}}},
  \bibinfo{author}{\bibfnamefont{A.~K.~Y.} \bibnamefont{{Li}}},
  \bibnamefont{and} \bibinfo{author}{\bibfnamefont{Y.}~\bibnamefont{{Chen}}},
  \bibinfo{journal}{Phys. Rev. D} \textbf{\bibinfo{volume}{103}},
  \bibinfo{eid}{104055} (\bibinfo{year}{2021}), \eprint{2101.08264}.

\bibitem[{\citenamefont{Ezquiaga et~al.}(2021)\citenamefont{Ezquiaga, Holz, Hu,
  Lagos, and Wald}}]{PhysRevD.103.064047}
\bibinfo{author}{\bibfnamefont{J.~M.} \bibnamefont{Ezquiaga}},
  \bibinfo{author}{\bibfnamefont{D.~E.} \bibnamefont{Holz}},
  \bibinfo{author}{\bibfnamefont{W.}~\bibnamefont{Hu}},
  \bibinfo{author}{\bibfnamefont{M.}~\bibnamefont{Lagos}}, \bibnamefont{and}
  \bibinfo{author}{\bibfnamefont{R.~M.} \bibnamefont{Wald}},
  \bibinfo{journal}{Phys. Rev. D} \textbf{\bibinfo{volume}{103}},
  \bibinfo{pages}{064047} (\bibinfo{year}{2021}),
  \urlprefix\url{https://link.aps.org/doi/10.1103/PhysRevD.103.064047}.

\bibitem[{\citenamefont{{Adhikari} et~al.}(2020)\citenamefont{{Adhikari},
  {Arai}, {Brooks}, {Wipf}, {Aguiar}, {Altin}, {Barr}, {Barsotti}, {Bassiri},
  {Bell} et~al.}}]{2020CQGra..37p5003A}
\bibinfo{author}{\bibfnamefont{R.~X.} \bibnamefont{{Adhikari}}},
  \bibinfo{author}{\bibfnamefont{K.}~\bibnamefont{{Arai}}},
  \bibinfo{author}{\bibfnamefont{A.~F.} \bibnamefont{{Brooks}}},
  \bibinfo{author}{\bibfnamefont{C.}~\bibnamefont{{Wipf}}},
  \bibinfo{author}{\bibfnamefont{O.}~\bibnamefont{{Aguiar}}},
  \bibinfo{author}{\bibfnamefont{P.}~\bibnamefont{{Altin}}},
  \bibinfo{author}{\bibfnamefont{B.}~\bibnamefont{{Barr}}},
  \bibinfo{author}{\bibfnamefont{L.}~\bibnamefont{{Barsotti}}},
  \bibinfo{author}{\bibfnamefont{R.}~\bibnamefont{{Bassiri}}},
  \bibinfo{author}{\bibfnamefont{A.}~\bibnamefont{{Bell}}},
  \bibnamefont{et~al.}, \bibinfo{journal}{Classical and Quantum Gravity}
  \textbf{\bibinfo{volume}{37}}, \bibinfo{eid}{165003} (\bibinfo{year}{2020}),
  \eprint{2001.11173}.

\bibitem[{\citenamefont{{Sathyaprakash}
  et~al.}(2011)\citenamefont{{Sathyaprakash}, {Abernathy}, {Acernese},
  {Amaro-Seoane}, {Andersson}, {Arun}, {Barone}, {Barr}, {Barsuglia}, {Beker}
  et~al.}}]{2011arXiv1108.1423S}
\bibinfo{author}{\bibfnamefont{B.}~\bibnamefont{{Sathyaprakash}}},
  \bibinfo{author}{\bibfnamefont{M.}~\bibnamefont{{Abernathy}}},
  \bibinfo{author}{\bibfnamefont{F.}~\bibnamefont{{Acernese}}},
  \bibinfo{author}{\bibfnamefont{P.}~\bibnamefont{{Amaro-Seoane}}},
  \bibinfo{author}{\bibfnamefont{N.}~\bibnamefont{{Andersson}}},
  \bibinfo{author}{\bibfnamefont{K.}~\bibnamefont{{Arun}}},
  \bibinfo{author}{\bibfnamefont{F.}~\bibnamefont{{Barone}}},
  \bibinfo{author}{\bibfnamefont{B.}~\bibnamefont{{Barr}}},
  \bibinfo{author}{\bibfnamefont{M.}~\bibnamefont{{Barsuglia}}},
  \bibinfo{author}{\bibfnamefont{M.}~\bibnamefont{{Beker}}},
  \bibnamefont{et~al.}, \bibinfo{journal}{arXiv e-prints}
  \bibinfo{eid}{arXiv:1108.1423} (\bibinfo{year}{2011}), \eprint{1108.1423}.

\bibitem[{\citenamefont{{Reitze} et~al.}(2019)\citenamefont{{Reitze},
  {Adhikari}, {Ballmer}, {Barish}, {Barsotti}, {Billingsley}, {Brown}, {Chen},
  {Coyne}, {Eisenstein} et~al.}}]{2019BAAS...51g..35R}
\bibinfo{author}{\bibfnamefont{D.}~\bibnamefont{{Reitze}}},
  \bibinfo{author}{\bibfnamefont{R.~X.} \bibnamefont{{Adhikari}}},
  \bibinfo{author}{\bibfnamefont{S.}~\bibnamefont{{Ballmer}}},
  \bibinfo{author}{\bibfnamefont{B.}~\bibnamefont{{Barish}}},
  \bibinfo{author}{\bibfnamefont{L.}~\bibnamefont{{Barsotti}}},
  \bibinfo{author}{\bibfnamefont{G.}~\bibnamefont{{Billingsley}}},
  \bibinfo{author}{\bibfnamefont{D.~A.} \bibnamefont{{Brown}}},
  \bibinfo{author}{\bibfnamefont{Y.}~\bibnamefont{{Chen}}},
  \bibinfo{author}{\bibfnamefont{D.}~\bibnamefont{{Coyne}}},
  \bibinfo{author}{\bibfnamefont{R.}~\bibnamefont{{Eisenstein}}},
  \bibnamefont{et~al.}, in \emph{\bibinfo{booktitle}{Bulletin of the American
  Astronomical Society}} (\bibinfo{year}{2019}), vol.~\bibinfo{volume}{51},
  p.~\bibinfo{pages}{35}, \eprint{1907.04833}.

\bibitem[{\citenamefont{Abbott et~al.}(2021)}]{LIGOScientific:2021izm}
\bibinfo{author}{\bibfnamefont{R.}~\bibnamefont{Abbott}} \bibnamefont{et~al.}
  (\bibinfo{collaboration}{LIGO Scientific Collaboration and Virgo
  Collaboration}), \bibinfo{journal}{Astrophys. J.}
  \textbf{\bibinfo{volume}{923}}, \bibinfo{pages}{14} (\bibinfo{year}{2021}),
  \eprint{2105.06384}.

\bibitem[{\citenamefont{Abbott et~al.}(2023)}]{LIGOScientific:2023bwz}
\bibinfo{author}{\bibfnamefont{R.}~\bibnamefont{Abbott}} \bibnamefont{et~al.}
  (\bibinfo{collaboration}{LIGO Scientific Collaboration, Virgo Collaboration
  and KAGRA Collaboration}), \bibinfo{journal}{arXiv e-prints}
  \bibinfo{eid}{arXiv:2304.08393} (\bibinfo{year}{2023}).

\bibitem[{\citenamefont{{G{\"u}nther}}(1965)}]{1965ArRMA..18..103G}
\bibinfo{author}{\bibfnamefont{P.}~\bibnamefont{{G{\"u}nther}}},
  \bibinfo{journal}{Archive for Rational Mechanics and Analysis}
  \textbf{\bibinfo{volume}{18}}, \bibinfo{pages}{103} (\bibinfo{year}{1965}).

\bibitem[{\citenamefont{{W{\"u}nsch}}(1990)}]{1990GReGr..22..843W}
\bibinfo{author}{\bibfnamefont{V.}~\bibnamefont{{W{\"u}nsch}}},
  \bibinfo{journal}{General Relativity and Gravitation}
  \textbf{\bibinfo{volume}{22}}, \bibinfo{pages}{843} (\bibinfo{year}{1990}).

\bibitem[{\citenamefont{McLenaghan}(1969)}]{mclenaghan_1969}
\bibinfo{author}{\bibfnamefont{R.~G.} \bibnamefont{McLenaghan}},
  \bibinfo{journal}{Mathematical Proceedings of the Cambridge Philosophical
  Society} \textbf{\bibinfo{volume}{65}}, \bibinfo{pages}{139–155}
  (\bibinfo{year}{1969}).

\bibitem[{\citenamefont{{Sonego} and {Faraoni}}(1992)}]{1992JMP....33..625S}
\bibinfo{author}{\bibfnamefont{S.}~\bibnamefont{{Sonego}}} \bibnamefont{and}
  \bibinfo{author}{\bibfnamefont{V.}~\bibnamefont{{Faraoni}}},
  \bibinfo{journal}{Journal of Mathematical Physics}
  \textbf{\bibinfo{volume}{33}}, \bibinfo{pages}{625} (\bibinfo{year}{1992}).

\bibitem[{\citenamefont{{Friedlander}}(1975)}]{1975weoc.book.....F}
\bibinfo{author}{\bibfnamefont{F.~G.} \bibnamefont{{Friedlander}}},
  \emph{\bibinfo{title}{{The wave equation on a curved space-time.}}}
  (\bibinfo{publisher}{Cambridge University Press, Cambridge},
  \bibinfo{year}{1975}).

\bibitem[{\citenamefont{{He} and {Wu}}(2022)}]{2022PhRvD.106l4037H}
\bibinfo{author}{\bibfnamefont{J.-h.} \bibnamefont{{He}}} \bibnamefont{and}
  \bibinfo{author}{\bibfnamefont{Z.}~\bibnamefont{{Wu}}},
  \bibinfo{journal}{Phys. Rev. D} \textbf{\bibinfo{volume}{106}},
  \bibinfo{eid}{124037} (\bibinfo{year}{2022}), \eprint{2208.01621}.

\bibitem[{\citenamefont{Aasi et~al.}(2015)}]{LIGOScientific:2014pky}
\bibinfo{author}{\bibfnamefont{J.}~\bibnamefont{Aasi}} \bibnamefont{et~al.}
  (\bibinfo{collaboration}{LIGO Scientific Collaboration}),
  \bibinfo{journal}{Class. Quant. Grav.} \textbf{\bibinfo{volume}{32}},
  \bibinfo{pages}{074001} (\bibinfo{year}{2015}), \eprint{1411.4547}.

\bibitem[{\citenamefont{He}(2021)}]{He:2021hhl}
\bibinfo{author}{\bibfnamefont{J.-H.} \bibnamefont{He}}, \bibinfo{journal}{Mon.
  Not. Roy. Astron. Soc.} \textbf{\bibinfo{volume}{506}}, \bibinfo{pages}{5278}
  (\bibinfo{year}{2021}), \eprint{2107.09800}.

\bibitem[{\citenamefont{Varma et~al.}(2019)\citenamefont{Varma, Field, Scheel,
  Blackman, Kidder, and Pfeiffer}}]{Varma:2018mmi}
\bibinfo{author}{\bibfnamefont{V.}~\bibnamefont{Varma}},
  \bibinfo{author}{\bibfnamefont{S.~E.} \bibnamefont{Field}},
  \bibinfo{author}{\bibfnamefont{M.~A.} \bibnamefont{Scheel}},
  \bibinfo{author}{\bibfnamefont{J.}~\bibnamefont{Blackman}},
  \bibinfo{author}{\bibfnamefont{L.~E.} \bibnamefont{Kidder}},
  \bibnamefont{and} \bibinfo{author}{\bibfnamefont{H.~P.}
  \bibnamefont{Pfeiffer}}, \bibinfo{journal}{Phys. Rev. D}
  \textbf{\bibinfo{volume}{99}}, \bibinfo{pages}{064045}
  (\bibinfo{year}{2019}), \eprint{1812.07865}.

\bibitem[{\citenamefont{{Ossokine} et~al.}(2020)\citenamefont{{Ossokine},
  {Buonanno}, {Marsat}, {Cotesta}, {Babak}, {Dietrich}, {Haas}, {Hinder},
  {Pfeiffer}, {P{\"u}rrer} et~al.}}]{2020PhRvD.102d4055O}
\bibinfo{author}{\bibfnamefont{S.}~\bibnamefont{{Ossokine}}},
  \bibinfo{author}{\bibfnamefont{A.}~\bibnamefont{{Buonanno}}},
  \bibinfo{author}{\bibfnamefont{S.}~\bibnamefont{{Marsat}}},
  \bibinfo{author}{\bibfnamefont{R.}~\bibnamefont{{Cotesta}}},
  \bibinfo{author}{\bibfnamefont{S.}~\bibnamefont{{Babak}}},
  \bibinfo{author}{\bibfnamefont{T.}~\bibnamefont{{Dietrich}}},
  \bibinfo{author}{\bibfnamefont{R.}~\bibnamefont{{Haas}}},
  \bibinfo{author}{\bibfnamefont{I.}~\bibnamefont{{Hinder}}},
  \bibinfo{author}{\bibfnamefont{H.~P.} \bibnamefont{{Pfeiffer}}},
  \bibinfo{author}{\bibfnamefont{M.}~\bibnamefont{{P{\"u}rrer}}},
  \bibnamefont{et~al.}, \bibinfo{journal}{Phys. Rev. D}
  \textbf{\bibinfo{volume}{102}}, \bibinfo{eid}{044055} (\bibinfo{year}{2020}),
  \eprint{2004.09442}.

\bibitem[{\citenamefont{Apostolatos}(1995)}]{PhysRevD.52.605}
\bibinfo{author}{\bibfnamefont{T.~A.} \bibnamefont{Apostolatos}},
  \bibinfo{journal}{Phys. Rev. D} \textbf{\bibinfo{volume}{52}},
  \bibinfo{pages}{605} (\bibinfo{year}{1995}).

\bibitem[{\citenamefont{Biwer et~al.}(2019)\citenamefont{Biwer, Capano, De,
  Cabero, Brown, Nitz, and Raymond}}]{Biwer:2018osg}
\bibinfo{author}{\bibfnamefont{C.~M.} \bibnamefont{Biwer}},
  \bibinfo{author}{\bibfnamefont{C.~D.} \bibnamefont{Capano}},
  \bibinfo{author}{\bibfnamefont{S.}~\bibnamefont{De}},
  \bibinfo{author}{\bibfnamefont{M.}~\bibnamefont{Cabero}},
  \bibinfo{author}{\bibfnamefont{D.~A.} \bibnamefont{Brown}},
  \bibinfo{author}{\bibfnamefont{A.~H.} \bibnamefont{Nitz}}, \bibnamefont{and}
  \bibinfo{author}{\bibfnamefont{V.}~\bibnamefont{Raymond}},
  \bibinfo{journal}{Publ. Astron. Soc. Pac.} \textbf{\bibinfo{volume}{131}},
  \bibinfo{pages}{024503} (\bibinfo{year}{2019}), \eprint{1807.10312}.

\bibitem[{\citenamefont{{Hannam} et~al.}(2022)\citenamefont{{Hannam}, {Hoy},
  {Thompson}, {Fairhurst}, {Raymond}, {Colleoni}, {Davis}, {Estell{\'e}s},
  {Haster}, {Helmling-Cornell} et~al.}}]{2022Natur.610..652H}
\bibinfo{author}{\bibfnamefont{M.}~\bibnamefont{{Hannam}}},
  \bibinfo{author}{\bibfnamefont{C.}~\bibnamefont{{Hoy}}},
  \bibinfo{author}{\bibfnamefont{J.~E.} \bibnamefont{{Thompson}}},
  \bibinfo{author}{\bibfnamefont{S.}~\bibnamefont{{Fairhurst}}},
  \bibinfo{author}{\bibfnamefont{V.}~\bibnamefont{{Raymond}}},
  \bibinfo{author}{\bibfnamefont{M.}~\bibnamefont{{Colleoni}}},
  \bibinfo{author}{\bibfnamefont{D.}~\bibnamefont{{Davis}}},
  \bibinfo{author}{\bibfnamefont{H.}~\bibnamefont{{Estell{\'e}s}}},
  \bibinfo{author}{\bibfnamefont{C.-J.} \bibnamefont{{Haster}}},
  \bibinfo{author}{\bibfnamefont{A.}~\bibnamefont{{Helmling-Cornell}}},
  \bibnamefont{et~al.}, \bibinfo{journal}{Nature}
  \textbf{\bibinfo{volume}{610}}, \bibinfo{pages}{652} (\bibinfo{year}{2022}),
  \eprint{2112.11300}.

\bibitem[{\citenamefont{Cornish et~al.}(2011)\citenamefont{Cornish, Sampson,
  Yunes, and Pretorius}}]{Cornish:2011ys}
\bibinfo{author}{\bibfnamefont{N.}~\bibnamefont{Cornish}},
  \bibinfo{author}{\bibfnamefont{L.}~\bibnamefont{Sampson}},
  \bibinfo{author}{\bibfnamefont{N.}~\bibnamefont{Yunes}}, \bibnamefont{and}
  \bibinfo{author}{\bibfnamefont{F.}~\bibnamefont{Pretorius}},
  \bibinfo{journal}{Phys. Rev. D} \textbf{\bibinfo{volume}{84}},
  \bibinfo{pages}{062003} (\bibinfo{year}{2011}), \eprint{1105.2088}.

\bibitem[{\citenamefont{Del~Pozzo et~al.}(2014)\citenamefont{Del~Pozzo, Grover,
  Mandel, and Vecchio}}]{DelPozzo:2014cla}
\bibinfo{author}{\bibfnamefont{W.}~\bibnamefont{Del~Pozzo}},
  \bibinfo{author}{\bibfnamefont{K.}~\bibnamefont{Grover}},
  \bibinfo{author}{\bibfnamefont{I.}~\bibnamefont{Mandel}}, \bibnamefont{and}
  \bibinfo{author}{\bibfnamefont{A.}~\bibnamefont{Vecchio}},
  \bibinfo{journal}{Class. Quant. Grav.} \textbf{\bibinfo{volume}{31}},
  \bibinfo{pages}{205006} (\bibinfo{year}{2014}), \eprint{1408.2356}.

\bibitem[{\citenamefont{{Chiba} and {Yokoyama}}(2017)}]{PBH_spin_2017}
\bibinfo{author}{\bibfnamefont{T.}~\bibnamefont{{Chiba}}} \bibnamefont{and}
  \bibinfo{author}{\bibfnamefont{S.}~\bibnamefont{{Yokoyama}}},
  \bibinfo{journal}{Progress of Theoretical and Experimental Physics}
  \textbf{\bibinfo{volume}{2017}}, \bibinfo{eid}{083E01}
  (\bibinfo{year}{2017}), \eprint{1704.06573}.

\end{thebibliography}

\end{document}